# Effect of Strand Longitudinal Thermal Conduction on Take-Off Properties of Cable-in- Conduit Superconductors

Alexander Anghel

*Abstract*—The effect of the strand longitudinal thermal conduction carried mainly by the stabilizing copper on the take-off (quench) behaviour of cable-in-conduit superconductors is investigated theoretically and numerically. An equal area criterion type of condition is found for the thermal equilibrium of the conductor. I show that the thermal conduction effect can be quantified in term of an effective, enhanced heat-exchange coefficient.

## I. INTRODUCTION

As the strand in a cable in conduit conductor (CICC) travels in and out of the high field region over one twist pitch of the cable, the local electric field and consequently the local strand temperature varies. The equilibrium strand temperature can be easily found by the method described in detail in [1]. In this calculation there is an important effect that has been neglected i.e. the effect of the longitudinal thermal conduction. Especially close to the quench point with very high electric field gradients at the peak field and for strands with large copper/non-copper ration it is expected that the thermal conduction along the strands would play an important role an surely cannot be neglected.

Based on physical intuition, one expects on general grounds that some of the heat produced in the high field region will be conducted away to the low field regions of the strand with the consequence that the local strand temperature will be lower in the high field region and higher in the low field region as compared to the values obtained using the simple model of [1] which neglects the longitudinal thermal conduction.

In this paper the condition of thermal equilibrium of a superconducting strand in a CICC under the effect of the longitudinal thermal conduction is investigated analytically and numerically using a simple 1D model of a power-law superconductor with a sinusoidal variation of the local magnetic field seen by the strand as it travels in the cable cross-section over a twist pitch length.

A. Anghel is with the Ecole Polytechnique Federale de Lausanne, CRPP-Fusion Technology, SULTAN facility, CH-5232 Villigen, Switzerland. (phone: +41 56 310 3723; fax: +41 56 310 3729; e-mail: anghel@psi.ch).

## II. EQUILIBRIUM CONDITION WITH FINITE LONGITUDINAL THERMAL CONDUCTION

In order to quantify the effect of longitudinal thermal conduction we adopt here a 1D model. We consider only the thermal conduction along a strand extended half of its twist pitch length based on the obvious symmetry. The basic equation is

$$\frac{\partial}{\partial s}\left(A_{cu}\kappa(T_{cond})\frac{\partial T_{cond}}{\partial s}\right) = H(T_{cond}, T_{he}) - G(T_{cond}, I_{op}) \quad (1)$$

with $s$, the coordinate along the strand (arclength) and $\kappa(T_{cond})$ the thermal conductivity of copper. The thermal conduction in the superconductor is an order of magnitude lower and can been neglected. Transversal effects, thermal and electrical are completely neglected here. The heating power $G$ is defined as

$$G(T_{cond}, I_{op}) = EI_{op} = E_c \left(\frac{I_{op}}{I_c}\right)^n I_{op} \quad (2)$$

where $E$ is the local electric field, $I_{op}$ the operating current and $I_c$ the critical current, a function of local magnetic field $B(s) = B_0 + kI_{op}\sin(2\pi s/p)$ and conductor temperature $T_{cond}$.
The cooling power $H$ is modeled by a simple convective term

$$H(T_{cond}, T_{he}) = hp_w(T_{cond} - T_{he}) \quad (3)$$

with $h$, some bare heat-exchange coefficient between strand and helium and $p_w$ the wetted perimeter of the strand.

The boundary conditions of this 1D equation emerge from the symmetry of the problem if we observe that both at the peak field and at the minimum field positions there is no heat flowing out or in, i.e. we have

$$\kappa(T_{cond})\frac{\partial T_{cond}}{\partial s}\bigg|_{s=0} = \kappa(T_{cond})\frac{\partial T_{cond}}{\partial s}\bigg|_{s=\frac{L_p}{2}} \quad (4)$$

where $L_p = \sqrt{p^2 + (2\pi r_c)^2}$ is the arclength of a strand corresponding to one twist pitch $p$ and $r_c$ is the cable radius.

Defining a new variable $U(T_{cond}) = \kappa(T_{cond})\frac{\partial T_{cond}}{\partial s}$, and eliminating the dependence on $s$ we can express Eq.(1) as

$$U(T_{cond})\frac{dT_{cond}}{dT} = \kappa(T_{cond})\left(H(T_{cond}) - G(T_{cond})\right) \quad (5)$$

Integrating this ordinary differential equation between $T_{\min} = T_{cond}(s=0)$ and $T_{\max} = T_{cond}\left(s = \frac{L_s}{2}\right)$, we get by assuming a constant heat conduction coefficient $\kappa(T_{cond}) = \kappa_0$

$$\frac{1}{2}\left(U^2(T_{\max}) - U^2(T_{\min})\right) = \kappa_0 \int_{T_{\min}}^{T_{\max}} \left(H(T) - G(T)\right) dT \quad (6)$$

In view of the condition expressed by Eq.(4), the condition $U(T_{\max}) = U(T_{\min})$ holds and the final result is

$$\int_{T_{\min}}^{T_{\max}} \left(H(T) - G(T)\right) dT = 0 \quad (7)$$

which is formally the same as the "equal area" criterion [2]. It shows that the stable equilibrium temperature profile along the strand, as it travels trough the field gradient pattern created by the self-field, is such that the equal area condition for heating and cooling is satisfied.

Although similar, this stable steady-state temperature profile satisfying the equal-area condition should not be confused with the minimum propagating zone (MPZ) profile because while the profile obtained here is a stable one, the MPZ is not. As opposed to the non-conductive case where the quench point appears when the condition of local thermal equilibrium at the peak field position is broken, in the conductive case the quench point is defined as the point where the equal area criterion does not hold anymore i.e. no "stable" longitudinal temperature distribution exists which is compatible with the equal area condition. At a given current if one starts to increase the helium temperature, a family of stable solutions is obtained when solving Eq.(1), until a certain helium temperature is reached for which there is no stable solution in the form of a steady state temperature profile. The last helium temperature compatible with a stable solution can be quoted as the quench helium temperature. For the conductor temperature at quench the conductor temperature at the peak field position can be used as a quench parameter. Similarly, in a current scan a fixed helium temperature is imposed and for growing current, stable solutions are obtained up to a quench current where there is no stable solution anymore.

Physically, it is obvious that the reason for the stable solution is that although there is an excess of heating over cooling in the high field region, this is conducted downwards to the low field region where there is an excess of cooling over heating which compensate the over heating in the high field region.

It can be shown (see below) that the net result is as if close to the peak field there is an enhancement of the heat exchange coefficient and in the low field region a reduction of the heat exchange coefficient.

In other words, if the longitudinal thermal conduction is taken into account, the "old" condition of local thermal equilibrium for each position along the strand $G(s) = H(s)$ is replaced by the more general integral condition Eq.(7), which allows locally for more heating than cooling providing somewhere else there is more cooling than heating.

The direct consequence is that under the same conditions a conductor with enhanced longitudinal thermal conduction (e.g. a larger copper/non-copper ratio) will quench at a higher temperature or current as the one with reduced thermal conduction effect or if experimental data are analyzed with the old (non-conductive) model, a high value of the heat exchange will result but a larger part of it is due to the longitudinal conduction effect.

### III. THE EFFECTIVE HEAT- EXCHANGE COEFFICIENT

The thermal conduction effect can be quantified by introducing an effective heat exchange coefficient. To this purpose, Eq. (1) can be rearranged such as to resemble the old equation $G = H$ used in the previous model [1]

$$G(T_{cond}, I_{op}) = H(T_{cond}, T_{he}) - \frac{\partial}{\partial s}\left(A_{cu}\kappa(T_{cond})\frac{\partial T_{cond}}{\partial s}\right) = \\ = h_{eq} p_w (T_{cond} - T_{he}) = H_{eq}(T_{cond}, T_{he}) \quad (8)$$

introducing an equivalent heat exchange coefficient defined by

$$h_{eq} = h - \frac{1}{p_w(T_{cond} - T_{he})}\frac{\partial Q}{\partial s} \\ Q = \kappa_{cu}(T_{cond}) A_{cu} \frac{\partial T_{cond}}{\partial s} \quad (9)$$

The numerical calculation shows (see below) that the equivalent heat exchange coefficient is indeed larger than the convective one in the high field region by a factor of 2.5 and considerably reduced in the low field region (almost zero).

### IV. NUMERICAL RESULTS: FINITE THERMAL CONDUCTION ALONG THE STRAND

In order to quantify the effect of the longitudinal thermal conduction, Eq. (1) has been solved numerically for a strand in a typical CONDOPT environment with the following parameters:

s/c=NbTi
power law index, n=15
convective heat exchange coefficient, h=400W/m2K

copper cross section, $S_{cu}$=1.5mm2
non-copper cross section $S_{nCu}$=0.2mm2
copper/non-copper=7.5
background field, $B_b$=5T
RRR=140
twist pitch=183mm

The copper resistivity and the thermal conduction coefficient κ were assumed temperature independent and corrected for the magneto-resistance effect. The critical current was calculated with the Bottura scaling [3].

Two cases were analyzed: one in which we completely neglected the longitudinal thermal conduction (κ=0) and the other one with thermal conduction fully taken into account (κ≠0). The first case corresponds largely also to the case with low copper/non-copper ratio (see below).

A first set of the results for an operating current of 230A and operating temperature (helium) of 4.8K are shown in Figs. 1 to 3. This combination is not a quench point for either of the two cases and is presented merely in order to have a common comparison basis for the conduction and no conduction cases. In this example calculation, the temperature profile along the strand over half of the twist pitch, the heating and cooling curves and the effective heat exchange coefficient calculated with Eq.(9) are compared for finite thermal conduction and no thermal conduction.

From Fig.1 (a) and (b) it can be seen that the conductor temperature at the peak field position (on the right side of the figure) is higher in the no-conduction case as in the conduction one. At minimum field position (on the left side of the figure) the conductor temperature in the conduction case is higher than the helium temperature while in the non-conduction case it is practically equal to the helium temperature. The temperature profile in the conductive case is flattened by the conduction. All the effects shown here seem to be small, but as the numerical calculation show, each tenth and sometimes hundredth of a K is important for this very sensible thermal equilibrium problem. The main reason is of course the power-law voltage-current characteristic.

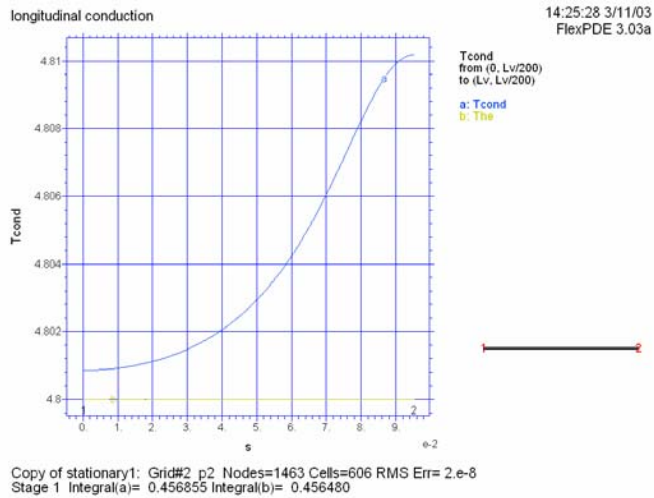

(a)

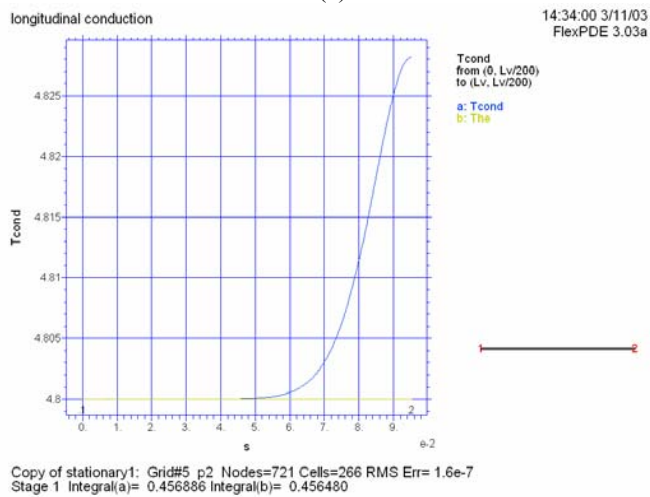

(b)

**Figure 1.** Conductor and He temperature along the strand from Bmin to Bmax (half twist pitch) with (a) finite thermal conduction and (b) no thermal conduction.

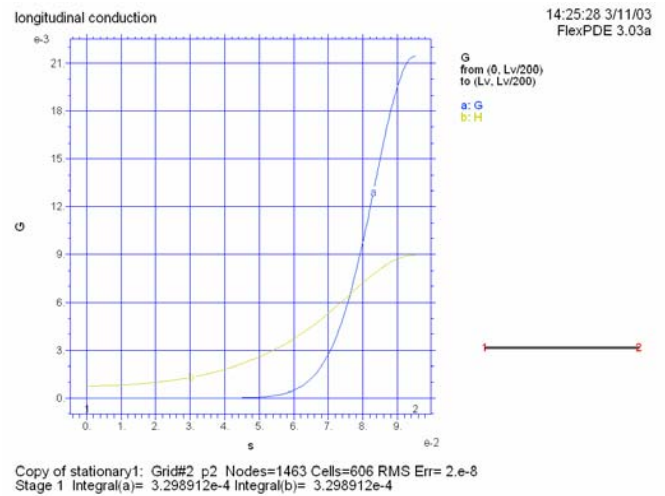

(a)

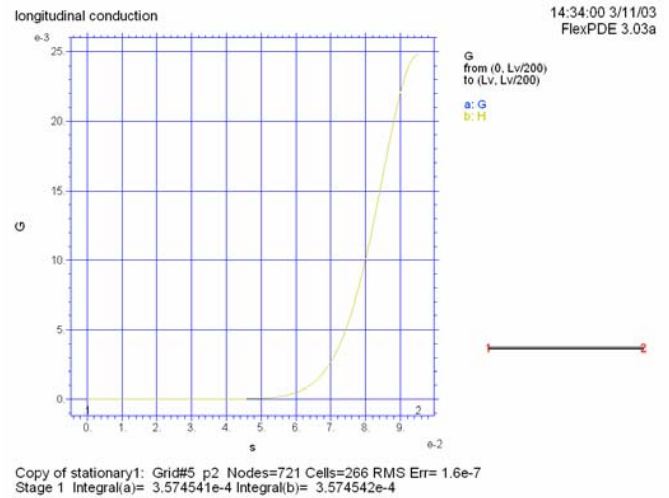

(b)

**Figure 2.** Index heating (G) and cooling (H) along the strand from Bmin to Bmax with (a) finite thermal conduction and (b) no thermal conduction.

of the strand as expected from the simplified model.

Fig.2 illustrates the heating and cooling along the strand length. In the conduction case (a) there is more cooling that heating over 75% of the strand length, starting from the minimum field position. In the high field region the situation reverses. In the non-conductive case heating and cooling are equal over the whole length. The fact that the equality $G = H$ does not hold for the conductive case but is replaced by the kind of 'equal area' condition can be checked by looking at the results of integration shown in the legend of Fig.2a (Integral(a)=Integral(b)=3.29891e-4). Oppositely, in the non-conductive case Fig.2b, $G \equiv H$ everywhere along the length.

The equivalent heat exchange coefficients for the two cases calculated with Eq.(9) are presented in Fig.3. In the conductive case Fig.3a, it is around 1000W/m2K (an enhancement factor of 2.5) at the peak field position and almost zero at $B_{min}$. In the non-conductive case, Fig.3b the equivalent heat exchange coefficient is of course the same as the convective one. The differences between the conductive and non-conductive cases at the quench point are presented in Table 1 and Table 2 showing the results for temperature and current scans. In the temperature scan case $I_{op}$=230A and the helium temperature is increased until a take-off takes place. In the current scan case $T_{he}$ is kept at 4.8K and the current is increased until a take-off occurs. All values are peak field values.

TABLE 1
Temperature scan: $I_{op}=I_q=230A$

| $\kappa=0$ | $\kappa \neq 0$ |
|---|---|
| $T_{q\_He}$=4.87K | $T_{q\_He}$=4.98K |
| $T_{q\_cond}$=4.946K | $T_{q\_cond}$=5.064K |
| $E_q/E_c$=28.986 | $E_q/E_c$=85.298 |
| $E_{avg}/E_c$=0.830 | $E_{avg}/E_c$=2.879 |
| $h_{eq}$=400W/m$^2$K | $h_{eq}$=1055W/m$^2$K |

TABLE 2
Current scan: $T_{he}=T_{q\_He}$=4.8K

| $\kappa=0$ | $\kappa \neq 0$ |
|---|---|
| $I_q$=236.4A | $I_q$=246A |
| $T_{q\_cond}$=4.9104K | $T_{q\_cond}$=4.9135K |
| $E_q/E_c$=41.08 | $E_q/E_c$=110.96 |
| $E_{avg}/E_c$=0.91 | $E_{avg}/E_c$=3.52 |
| $h_{eq}$=400W/m$^2$K | $h_{eq}$=1084W/m$^2$K |

It is clear from Tables 1 and 2 that what seems to be a premature quench for $\kappa=0$ ($E_{avg}<E_c$ at take-off) it is a normal quench for $\kappa \neq 0$ i.e. $E_{avg}>E_c$ at take-off. Also simulation with the non-conductive model with $h=h_{eq}$ give the same result for the quench point as the conductive model.

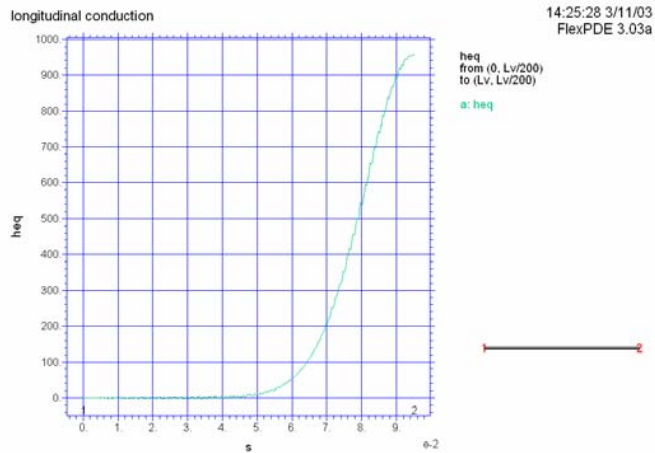

(a)

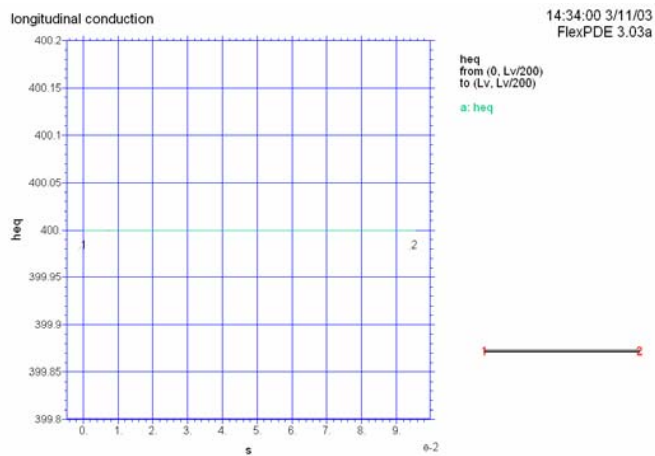

(b)

**Figure 3.** Variation of the equivalent heat exchange coefficient along the strand from Bmin to Bmax with finite thermal conduction (a) and no thermal conduction (b).

This points suggest that the experimental data can still be analysed with the old model (neglecting conduction) if the enhanced heat exchange coefficient is used. Alternatively, if the heat exchange coefficient is used as a fit parameter (the usual procedure), the value of $h$ should be reduced by a factor between 2 and 3 in order to get the real value of the heat exchange coefficient.

In another series of numerical simulations I have investigated in detail the effect of varying the copper cross-section. The results indicate that up to a copper/non-copper ratio of 1 the results are very close to the $\kappa=0$ case. The effect of longitudinal thermal conduction is therefore an important effect only for cables made of strands with copper/non-copper ratios greater than 1.

For conductors with segregated copper, even if not all of the external copper could be assessed to effectively participate to the longitudinal conduction, the effective copper/non-copper ratio can easily exceed 1.

A typical dependence of the effective heat exchange coefficient and of quench current as a function of the copper/non-copper ratio is presented in Fig.4.

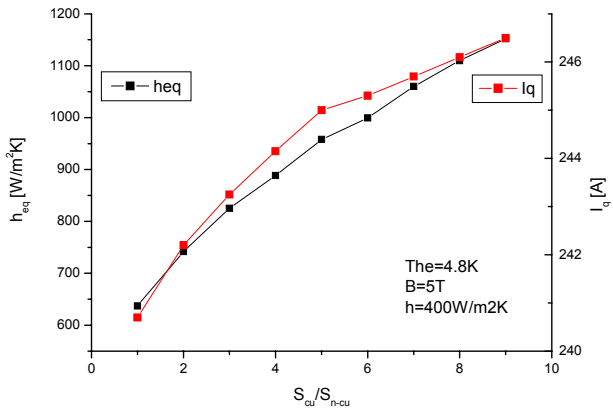

**Figure 4.** Variation of the effective heat-exchange coefficient and quench current with the copper/non-copper ratio

TABLE 3
Temperature scan with current redistribution: for $I_{op}=I_q=230A$

| $\kappa=0$ | $\kappa\neq0$ |
|---|---|
| $T_{q\_He}$=4.89K | $T_{q\_He}$=5.01K |
| $T_{q\_cond}$=4.986K | $T_{q\_cond}$=5.112K |
| $E_q/E_c$=36.63 | $E_q/E_c$=97.71 |
| $E_{avg}/E_c$=0.976 | $E_{avg}/E_c$=3.76 |
| $h_{eq}$=400W/m$^2$K | $h_{eq}$=1000W/m$^2$K |

From a log-log plot the following dependence could be extracted

$$h_{eq} \simeq 621 \left(\frac{S_{cu}}{S_{n-cu}}\right)^{0.272}$$

It is known that for those copper/non-copper ratios where a significant contribution of longitudinal conduction is expected, also the current redistribution between copper and s/c filaments play a role. With more copper cross section the current from the s/c is transferred to copper at earlier stages i.e. at lower electric fields during a temperature or current scan. This effect is illustrated in Table3 for a temperature scan with the same parameters as in Table 1 but now with the current transfer modelled with a parallel resistor model. Inclusion of current transfer is only marginally efficient in reducing the effective heat exchange coefficient

V. CONCLUSION

The effect of the longitudinal thermal conduction on the take-off behaviour of a conductor has been investigated.

It is shown that, in contrast to the non-conductive case where the equilibrium temperature of the strand is a local effect and a result of the local balance between the heat generation and cooling, in the conductive case the thermal equilibrium is the result of a more general condition expressed by a kind of "equal area" criterion.

Along the strand path a stable temperature profile is established which is compatible with the magnetic field distribution along the strand. This condition holds for all operating conditions up to a current (or temperature) where the stability of the solution disappears and this defines the take-off (quench) point.

It was shown that the thermal conduction effect can be described by a position dependent effective heat-exchange coefficient.

Numerical simulations with the 1D model have shown that the effect of thermal conduction is important for copper/non-copper ratios above 1. It was shown that due to the longitudinal conduction the effective heat-exchange coefficient is variable along the strand with an important enhancement by a factor of 2.5 at the peak-field position. This explains the overestimated heat exchange coefficients obtained usually when analyzing experimental data without taking the longitudinal thermal conduction into account [4].

A first check with the combined longitudinal conduction and current redistribution between copper and s/c has been also done showing a certain influence. In the case analyzed this is small but I cannot exclude that there could be cases where the effect could be important.

Only the internal copper was used in the present investigation. Segregated stabilizing copper in the cable will create an alternative, parallel path for heat conduction. Although it is not clear at the present time if the segregated copper should be included or not in the calculation, it is obvious that if it is to be included then the copper/non-copper ratio of cables with segregated copper can easily exceed 1 and the thermal conduction effect is indeed important.

This study is by no means exhaustive. Other regions of the parameter space: background fields, temperatures and/or other conductor designs have not yet been investigated and a more systematic study is necessary.